\documentclass[12pt]{article}

\usepackage[utf8]{inputenc}
\usepackage[english]{babel}
\usepackage{color, graphicx}

\usepackage[dvipsnames]{xcolor}
\usepackage[pdftex,%
plainpages=false,%
linktocpage=true,%
a4paper=true,%
colorlinks=true,%
linkcolor=NavyBlue,%
citecolor=BrickRed,%
urlcolor=NavyBlue,%
bookmarks=true]{hyperref}

\title{Cloudifying the Curriculum with AWS}
\author{Michael Soltys\footnote{California State University Channel
Islands, Professor in the Department of Computer Science,
URL: \href{http://www.msoltys.com}{www.msoltys.com}, Email:
\href{mailto:michael.soltys@csuci.edu}{michael.soltys@csuci.edu}}}
\date{\today}

\begin{document}
\maketitle

\begin{abstract}
The Cloud has become a principal paradigm of computing in the last ten
years, and Computer Science curricula must be updated to reflect that
reality. This paper examines simple ways to accomplish curriculum
cloudification using Amazon Web Services (AWS), for Computer Science
and other disciplines such as Business, Communication and Mathematics.
\end{abstract}

\section{Introduction}

Whether aware of it or not, most computer users have moved to the
Cloud in the last fifteen years. They have done so by using Webmail,
Google Docs, photo-sharing services or on-line gaming. Business has
followed the trend a few years later, by replacing or supplementing
in-house data centers with cloud services. 

At Universities, Computer Science (CS) and Information Technology (IT)
faculty have been using the Cloud just like everyone else, but many
have realized over the last decade that they can access compute power
in the cloud without making large capital investments on campus, and
can start using services, such as AWS virtual computers, with ease and
speed.  Since faculty work on research projects with students, senior
students had to acquire cloud skills as well.

At the same time, job sites such as LinkedIn listed knowledge of the
Cloud as the top job skill over the last five years
\cite{linkedin-skills-2019}, and currently 14\%\ of {\em all} job
listings require some understanding of the fundamentals of the cloud.
As students and their families invest in costly education, they are
keenly aware of this job market reality \cite{forbes-2019} (see blog
entry \cite{my-blog-post-5123}), and especially of AWS' growing
dominance in this field \cite{my-blog-post-5224}.  Therefore, a demand
for cloud instruction arose on campuses, and some faculty started
adding cloud content to the curriculum.  One of the earliest adopters
of a cloud curriculum was the Santa Monica College
\cite{santa-monica-2018}.

This paper examines methods to cloudify the curriculum with AWS
\cite{aws-overview-oct2019}. It starts from the assumption that the CS
and IT curricula are well formed and mature, and that radical changes
are not desirable. Instead, non-invasive and easy to implement ideas
are proposed. Furthermore, the paper examines how cloud content can be
presented to students in other disciplines, in particular students in
Business, Communication and Mathematics. We concentrate on four-year
Universities, while we are aware the two-year Community Colleges often
led the charge in cloudification.

A few years ago we started Engineering on our campus at California
State University Channel Islands, with Mechatronics Engineering. We
found it very helpful to discuss with other Universities their
experience in starting Mechatronics, and we found it especially
helpful to consult the experience of the University of Utah which was
recorded in \cite{meek-2002}. The goal of this paper is to provide a
similar template but for cloud adoption. The paper is meant to offer
suggestions, rather than definitive solutions, and represents the
opinions of the author. Each institution should adopt the cloud
according to its particular circumstances.

Feedback to the author from those who have experience in bringing the
Cloud to the curriculum would be most welcome.

\section{Cloudifying the Curriculum}

At the beginning of the discussion on curriculum cloudification we
decided to examine the students we intended to serve, and we grouped
them as follows:

\begin{enumerate}
\item Computer Science and Information Technology undergraduate
majors.
\item Masters in Computer Science.
\item Business, Communication and Mathematics majors.
\item Working professionals.
\end{enumerate}

As these students are served by different curricular pathways, this
allowed us a divide-and-conquer approach to cloudification. And by
``serve'' we mean that we expose them to the Cloud to the degree they
want, so that they can take advantage of the cloud-favorable job
market.

\subsection{CS and IT}

The first group, CS and IT students, form our largest contingent of
students, about five hundred currently, but growing quickly, as we
have doubled our number of such majors over the last three years. 
About 80\%\ of those students are in CS.

\subsubsection{CS}\label{sec:cs}

Our CS curriculum is based on the ACM curriculum 2013
(\cite{cur2013}), where the four main pillars of AWS, {\em Compute},
{\em Storage}, {\em Databases} and {\em Networking}, are covered in
depth. For example, Compute and Storage are covered in a sequence of
three ``systems'' classes, COMP 162, 262 and 362, that cover
everything from computer architecture and assembler to advanced topics
in operating and file systems.  

Databases are covered in our senior course COMP 420, and Networking in
our senior course COMP 429. CS students also receive a solid grounding
in programming capped with a junior Software Engineering class COMP
350. 

Thus, our senior CS students are conceptually more than ready to study
for the AWS Solutions Architect certification, and in fact any of the
AWS certifications, except the professional one that requires some
years of practical experience.

A light-footed approach to deliver cloud content to CS majors is to
``sprinkle'' it throughout our undergraduate classes as use cases and
illustrations of concepts being taught.  This approach {\bf relies on the
faculty discretion to decide on examples and platforms}\footnote{See
\cite{academic-freedom-2010} for a discussion of the meaning of
``Academic Freedom.'' Keep in mind that an instructor is responsible
for covering the content in the syllabus of a course, as the syllabus
is a contract between the university and the student. However, point
11 of \cite{academic-freedom-2010} says that ``Academic freedom gives
faculty members substantial latitude in deciding how to teach the
courses for which they are responsible.'' In our case, instructors can
illustrate the concepts with examples, use cases and technologies of
their choosing, especially since syllabi tend to be technology
agnostic.}, and does not require time-consuming
curricular changes and
paying attention to accreditation issues with
ABET\footnote{\url{https://www.abet.org/}} and
WASC\footnote{\url{https://www.acswasc.org/}}. Core concepts are
presented in class as promised in the class syllabus, but the
illustration of those concepts with AWS tools and use cases falls
under the discretion of the instructor.

Some instructors may choose to use AWS, some may not. However, for all
students interested in absorbing more cloud material we offer two ways
to do so: one, to use AWS in their capstone (a full year class in
senior year), two, to take a special topics COMP 490 class that
follows the AWS Solutions Architect certification. Both ways may be
chosen.

Of course, as cloud content becomes embedded in the curriculum in the
upcoming years, reflecting the new paradigm in the industry, some
instructors may choose to formalize it in the syllabi, possibly
requiring curriculum committee approvals; however, this does not
preclude teaching the material already now under our ``faculty
discretion'' model explained above.

\subsubsection{IT}

In our school the IT program was brought about in order to accommodate
transfer students from local colleges. The program was built using a
grant, and at the time of its founding it was designed to emphasize
web development.  The principal difference between CS and IT is that
IT students take less mathematics classes, less programming, and have
more electives (partly to assist in the transfer that would allow
their community college classes to count toward graduation). 

In recent years we have been looking for a new, more current, emphasis
for our IT program, and we considered business or cybersecurity. But
we started discussing the possibility to make it Cloud centric by
adding AWS to the curriculum more deliberately.  This would require
better coordination with Community Colleges.  However, at this moment,
the approach to offering AWS to our IT students is similar to CS: the
choice to take an AWS special topics class in the senior year. 

\subsection{Masters in CS}

In the Masters program in Computer Science, we introduced the cloud
into three core classes: Networking, Security and Software
Engineering.

\subsubsection{Networking}

For the last decade we offered a graduate course in Networking with
the title ``Cloud Computing'' (COMP 529) \cite{my-page-4255}. This
course is a graduate version of our undergraduate Networking class
(COMP 429). Recently we introduced the material from the AWS Solutions
Architect (SA) certification into its syllabus. The symbiosis of
Networking and AWS is a very natural way to deliver this class, and
students get an offering that not only covers the entire SA
certification, but goes far beyond it. For example, students learn
about routing at the IP level, and flow and congestion algorithms at
the TCP level, as well as most of the content in \cite{kurose}.

\subsubsection{Security}

Cybersecurity is one of the areas of emphasis in our department, and
we have been teaching a graduate version of Cybersecurity (COMP 524)
for several years \cite{my-page-3505}. (The undergraduate version of
that course is COMP 424.) Since the author received an AWS Specialty
Security (SS) certification in December 2019, we have been updating
the class material to include the entire certification curriculum as
use cases and examples.  The author wrote a manuscript containing the
notes for the class, with advanced material such as Cryptography,
Malware, Distributed Denial of Service attacks, as well as an overview
of most of the tools in Kali Linux.

Adding the material in the AWS SS certification allows us to update
the course to a more current offering as it now includes security in
the Cloud.  This includes details of Identity Access Management (IAM),
S3, Security Policies, Logging and Monitoring, Key Management System
(KMS) and use cases of CloudTrail, CloudWatch, Inspector, Cloud
Formation, Cloud Config, and other AWS tools that allow a more
hands-on approach.

The class has really improved as the result of including the AWS SS
material. For example, it is one thing to talk about the importance of
data encryption at rest for PCI compliance; it is quite another to
demonstrate to the students the automated enforcement of such
compliance with Amazon Macie machine learning tool.  The upgraded
class will be taught for the first time in the fall 2020.

\subsubsection{Software Engineering}

We invested quite heavily in our Software Engineering offering, both
at the undergraduate and graduate levels. A significant portion of our
students are employed as programmers, and we want them to be team
leaders and to think as engineers in their future roles in the
industry.  Programming and Software Engineering skills are in great
demand by employers.

Hence, we invest effort in teaching Software Engineering, using the
{\em Agile Methodology}, and standard engineering techniques such as
design, requirements, specifications, etc. Students learn to work on a
team, using collaborative tools such as GitHub.

We have taught the graduate version of Software Engineering, COMP 550,
using AWS as a platform. There were two large teams of students: the
first team built a collaborative code editor similar in spirit to
Google Docs, but with text highlighting for a few languages. The
second team used Unity to develop a network based game.  Both use AWS
to host their application and data storage of user information.

The plan for the near future is for some faculty to obtain the AWS
Developer certification, and combine the curriculum in that
certification with our own very well developed content.

\subsection{Business, Communication and Mathematics}

It is easy to include Mathematics majors in our cloudification
initiative, as CS and Math are closely integrated as departments, and
our majors are not far from being double-majors. Hence what we wrote
about CS students in Section~\ref{sec:cs} applies directly to Math
students. Mathematics also delivers emphases in Data Analytics and
Imaging, both of which can take advantage of AWS tools such as Amazon
Athena, Redshift, Rekognition, etc.  We plan to introduce these tools
into Mathematics.

As was mentioned in the Introduction, LinkedIn has listed the Cloud as
a top job skill over the last five years (in 2020 the top spot may be
replaced by Blockchain \cite{linkedin-skills-2020}, but even so the
Cloud will be a close second).  However, the top {\em soft skills}
include persuasion, collaboration, adaptability and emotional
intelligence, associated with Business and Communication
students. As mentioned in the Introduction, 14\%\ of jobs
will require some cloud knowledge, and so Business and Communication
students can expect to work in environments where the Cloud will be
central to a business mission.

We have designed a new course, {\em Online Communication and Society},
COMP 347, cross-listed with Communication, where we have embedded the
Cloud with Business and Communication students in mind
\cite{my-page-4527}. In this class,
the students learn the basics of AWS, launch an EC2 instance, and
install a LAMP stack on it in order to install Wordpress. Once
Wordpress is running, they integrate it with social media (e.g., LinkedIn
and Twitter), and publish content as posts and pages. 

In order to make the exercise more practical, we propose to the
students as framework the following use case: {\em you have been hired
by a non-profit with few resources but big ambitions, and you need to
promote your mission with a minimal budget}. The content was developed
by the author, and the technical material was based on AWS excellent
whitepapers, in particular \cite{aws-tutorial-lamp,
aws-tutorial-wordpress}.

It is important to remember that for most companies, the decision to
move to the cloud is principally a business decision, not an IT
decision. Both the basic AWS Cloud Foundations certification, and the
advanced AWS Solutions Architect certification emphasize the business
aspect of cloud solutions. This is a natural arena for business
students, and as the CEO of AWS mentioned at 2019 re:Invent in Las
Vegas, only 3\%\ of world IT is currently in the cloud (while it
comprises over 60\%\ of all IT spending!), and so as more IT moves to
the cloud, this will be an area of activity for business oriented
students.

One more thing ought to be mentioned: we have joined the {\em AWS
Activate for Startups} program, where we can nominate startups for low
cost AWS credits. This works well with our campus entrepreneurship
initiative.

\subsection{Working professionals}

We are offering two certification classes open to non-matriculated
students (i.e., anyone who wants to take them): AWS Cloud Foundations
and AWS Solutions Architect, both of them through AWS Academy
\cite{my-blog-post-5203}.  These
classes are aimed at working professionals, they are eight weeks long,
and delivered mostly online. We will seek approval for a Cloud
Certificate through our campus curriculum committee. The mechanism for
delivering this certificate program is our Extended
University\footnote{Extended University is often called the ``School
of Continued Education,'' but diplomas from EU are simply diplomas
from our school. It is an internal administrative entity, but part of
the university as such.}, which does not receive State funding, and is
therefore more flexible in how it can mount programs.

The Cloud certificate is a service to our community, especially the
Navy (there are two large Navy bases in Ventura County), as well as
the industry comprised of the Department of Defense contractors as
well as companies in the ``101 Technology Corridor.'' Furthermore, as
\cite{conley-2019} writes in his paper {\em The great enrollment
crash}, most universities are expecting up to a 15\%\ drop in
enrollment due to demographics over the next decade, and serving the
working population will become only more important to our viability as
an institution.

\section{Ancillary Efforts}

{\bf Instructor training and certification:} there has to be a
critical mass of instructors, both tenure-track faculty and lecturers,
who are willing to learn and deliver the AWS content. We currently
have two AWS Academy accredited instructors, and three more in the
pipeline. The author is the principal point of contact for AWS Academy
on campus, and can nominate interested educators.

{\bf Communicate to the faculty the benefits of cloudifying:}
tenure-track faculty have invested many years into their careers: a
PhD, then post-docs, then working toward tenure, this can be 15 years
or more. At the same time, in an era of great specialization, they
work to become respected members of a research community. Their time to
learn the foundations of other areas is very limited. One way to
motivate them is to present the advantages of AWS in their research;
this is one of the aims of the AWS Ambassador program.

{\bf Curriculum Changes vs Faculty Discretion:} As already discussed
in the paper in Section~\ref{sec:cs}, it may be more effective at
first to invoke faculty discretion in introducing cloud content rather
than an effort to codify a new curriculum.  Curricular decisions are
always lengthy, take up several teaching cycles, and the department
may end up missing the boat of cloudification. We argue that it is
best to cloudify {\em now} with faculty discretion, while pursuing
curricular changes as they become necessary. This provides a mechanism
to synchronize the different speeds of innovation in industry versus
the academia.

{\bf Advisory Board:} a departmental advisory board can be a great
ally in bringing about the cloudification of the curriculum. For one,
the members of the board are probably contemplating a possible move to
the cloud, and understand the need to train the workforce. In our case
we are lucky to have a large and supportive advisory board, comprised
of about 20 local companies, and we kept them abreast of the AWS
initiative from the beginning. \cite{my-url-advisory-board}

{\bf Support of the administration:} faculty have ten great
ideas every day, but they all require precious and limited campus
resources. It is imperative to have the support of the administration
while pursuing this initiative, which, as in the case of faculty,
requires communicating the benefits and low costs of the effort. In
our case we were very fortunate to have such support. It helped to
receive a \${35}K credit grant from AWS.

\section{Conclusion}

Cloud Computing is a new paradigm and also an inversion of an old one.
When the telegraph was invented in the early 1800s, it was ``smart at
the edges'' (the unit and its operator), while the connecting network
was very simple (just a cable and repeaters \cite{petzold-2000}). Then
came the telephone, with dumb terminals (rotary phones) at the edges,
and a complex network inside with switchboards for circuit-switching.
Then the paradigm was inverted once again with the Internet, where now
we had smart terminals at the edges, and a relatively simple
packet-switching network inside. The simplicity of the network allowed
for the quick innovation that followed, and the result of that
innovation was to invert the paradigm once again with Cloud Computing,
where now computers at the edge are simply entry portals into a
complex network, i.e, ``the Cloud'' \cite{Castro_2019}.

Computer Science is a field with a fast paced proposal of new
paradigms, and curricula must, on the one hand, offer the fundamentals
of a now well establish field, but on the other hand be nimble enough
to accommodate the fast rate of innovation. One way to accomplish this
is to offer a core of fundamentals in each area, illustrated with use
cases from the latest trends.  We propose that this is a practical and
effective approach to introduce Cloud Computing into the curriculum.

\section{Acknowledgments}

We are grateful to the {\em AWS Academy} and {\em AWS Educate} teams
for guiding us through the process, and for all high quality
educational materials they offer for free to participating
institutions. We are also grateful to Michael Berman for comments on
an early draft.



\end{document}